\documentclass[aps,twocolumn]{revtex4}
\usepackage{amsfonts}
\usepackage{amsmath}
\usepackage{amssymb}
\usepackage{graphicx}

\begin{document}
\title[The meta book]{The meta book and size-dependent properties of written language}
\author{Sebastian Bernhardsson, Luis Enrique Correa da Rocha and Petter Minnhagen}
\affiliation{Dept. of Physics, Ume\aa \ University. 901 87 Ume\aa . Sweden}
\pacs{PACS number}

\begin{abstract}
Evidence is given for a systematic text-length dependence of the power-law index $\gamma$ of a single book. The estimated $\gamma$ 
values are consistent with a monotonic decrease from 2 to 1 with increasing length of a text. A direct connection to an extended Heap's law
is explored. The infinite book limit is, as a consequence, proposed to be given by $\gamma = 1$ instead of the value $\gamma=2$ expected if the Zipf's law was ubiquitously applicable. In addition we explore the idea that the systematic text-length dependence can be described by a meta book concept, which is an abstract representation reflecting the word-frequency structure of a text. According to this concept the word-frequency distribution of a text, with a certain length written by a single author, has the same characteristics as a text of the same length pulled out from an imaginary complete infinite corpus written by the same author.
\end{abstract}

\maketitle

\section{Introduction}
The development of the spoken and written language is one of the major transitions in evolution \cite{Maynard}. It has given us the advantage to easily and efficiently transfer information between individuals and even between generations. It could be argued that it is clear why language was evolved in general, but it is harder to explain the reason for its structure. The structure of language has been studied as early as the Iron age in India and is still, to this day, a popular subject. 

The field had a boost after George Kingsley Zipf, around 75 years ago, found an empirical law (Zipf's law) \cite{Zipf32} describing a seemingly universal property of the written language. It states that the number of occurrences of a word in a long enough written text falls off as $1/r$ where $r$ is the occurrence-rank of a word (the smaller rank, the more occurrences) \cite{Zipf32} \cite{Zipf35} \cite{Zipf49} \cite{mitzenmacher03} \cite{baayen01}. 
This in turn means that the normalized word-frequency distribution (wfd) follows the expression $P(k) \propto 1/k^2$, where $P(k)$ is the probability to find a word which appears $k$ times in a text \cite{baayen01}. This empirical law is generally believed to represent some ubiquitous nature of the wfd, and has inspired the development of several models reproducing this structure \cite{simon55}\cite{mandelbrot53}. However, empirically one typically finds that the wfd follows a power-law distribution with an exponent smaller than 2 \cite{Cancho}\cite{Bernhardsson09}.
It was also reported in Ref.\ \cite{Bernhardsson09} that the exponent (commonly denoted as $\gamma$) for a power-law description of the wfd seems to change with the length of a text, rather than being constant. 

Another property is the number of different (unique) words, $N$, as a function of the total number of words in a book, $M$ (In this context a book is a sequence of words where words are defined as collections of letters separated by spaces).
The conventional way of describing this relation is by using Heap's law \cite{heaps}, which states that $N \propto M^{\alpha}$, where $0 < \alpha < 1$ is a constant. 

In this paper we present, and give evidence for, a meta book concept which is an abstract picture of how an author writes a text. 
We suggest a systematic text-length dependence for the wfd which is directly connected to an extended Heap's law with an $\alpha$ changing from 1 to 0 as the text length is increased from $M=1$ to infinity.

\section{The meta book concept}
We start by studying the above mentioned property, $N(M)$.
Figure 1 shows this curve for three different authors (\textbf{H}ardy, \textbf{M}elville and \textbf{L}awrence). We have created very large books by attaching novels together, in order to extend the range of book sizes (see appendix A for a full list of books). The curve shows a decreasing rate of adding new words which means that $N$ grows slower than linear ($\alpha<1$) \cite{baayen01}\cite{Bernhardsson09}. Also, for a real book, $N(M=1)=1$ which means that the proportionality constant in Heap's law must be one. So, if $N = M^{\alpha}$ then $\alpha = \ln N/\ln M$. This quantity is plotted in the inset of Fig.\ 1 and the data shows that $\alpha$ is decreasing as a function of the size, ruling out the possibility to accurately describe the $N(M)$-curve using a constant $\alpha$. A plausible scenario would be that $\alpha$ continue to decrease assymptotically towards zero as $M$ reaches infinity. This would mean that the $N(M)$-curve saturates and $N(M\rightarrow \infty)$ is finite.

\begin{figure}[!t]
\begin{center}
 \includegraphics[width=\columnwidth]{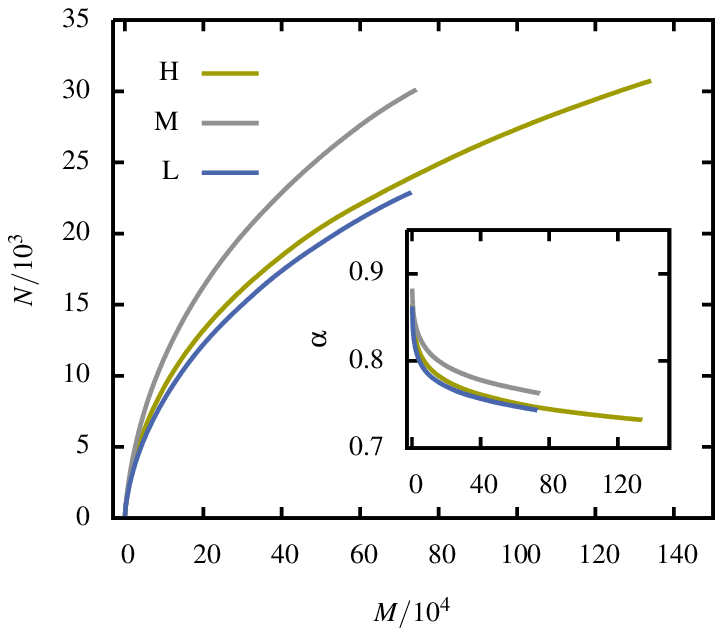}
\end{center}
\caption{The number of different words, $N$, as a function of the total number of words, $M$, for the authors \textbf{H}ardy, \textbf{M}elville and \textbf{L}awrence. The data represents a collection of books by each author. The inset shows the exponent $\alpha=\ln N/\ln M$ as a function of $M$ for each author.}
\end{figure}

When the length of a text is increased, the number of different words is also increased. However, the average usage of a specific word is not constant, but increases as well. That is, we tend to repeat the words more when writing a longer text. One might argue that this is because we have a limited vocabulary and when writing more words the probability to repeat an old word increases. But, at the same time, a contradictory argument could be that the scenery and plot, described for example in a novel, are often broader in a longer text, leading to a wider use of ones vocabulary. There is probably some truth in both statements but the empirical data seem to suggest that the dependence of $N$ on $M$ reflects a more general property of an authors language.

\begin{figure}[!t]
\begin{center}
 \includegraphics[width=\columnwidth]{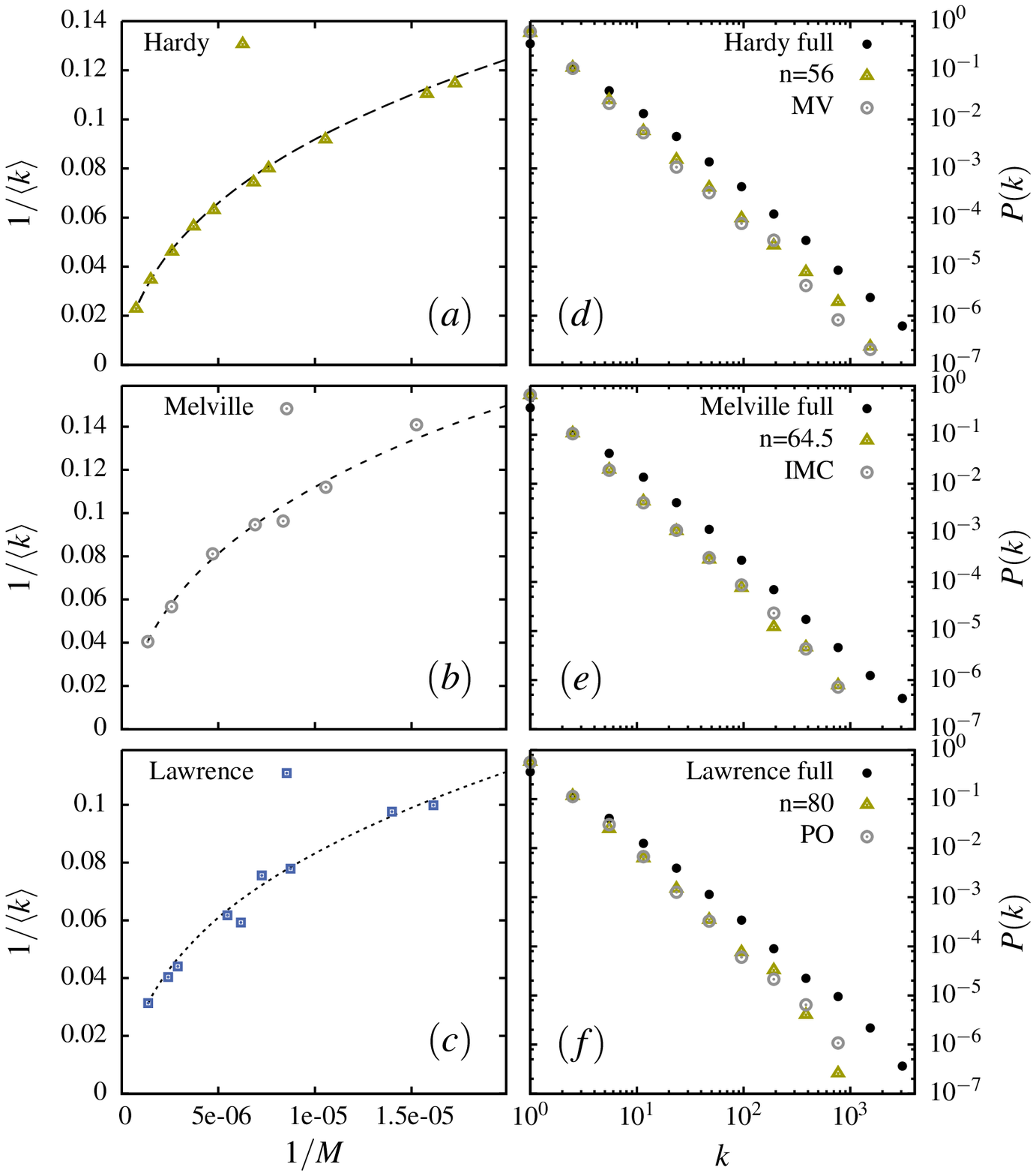}
\end{center}
\caption{Evidence in favor of the meta book concept: (a)-(c) The average frequency for a word as a function of the size of the book ($M$) plotted as $1/\langle k \rangle(\frac{1}{M})$ for the three authors. The long dashed, short dashed and dotted curves correspond to the $N(M)$-curve as $N/M(\frac{1}{M})$ for the biggest collection of books by each respective author. The $\langle k \rangle$ for a small book is close to the same as for a section (of the same size) of the bigger book.
(d)-(f) The word frequency distribution for an $n$th-part (triangles) of the full collection of books (filled circles) compared to a small book (open circles) of the same size as the $n$th-part. The wfd is approximately the same for a small book as for a section (of the same size) of a big book.}
\end{figure}

For every size of a text, the average occurrence for a word can be calculated as $\langle k \rangle = M/N$. This means that the $N(M)$-curve can be converted into a curve for the average frequency as $\frac{M}{N}(M) = \langle k \rangle(M)$. This curve is shown in Fig.\ 2a-c for the three different authors. Each point represent a real book or a collection of books and the curves represent the $\langle k \rangle(M)$-curve for the full collection of books for each author (i.e.\ same data as in Fig.\ 1). The data is plotted as $1/\langle k \rangle$ as a function of $1/M$ in order to get a feeling for the asymptotic behavior as $M$ reaches infinity.

The overlap between the line and the points means that the average frequency of a word (and consequently also $N$) in a short story is to good approximation the same as for a section, of equal length, from a larger text written by the same author.
Note that the texts has to be written by the same author since the overlap would not be nearly as good if books by Lawrence were compared to the curve by Melville.

In Fig.\ 2d-f, we literally pull out sections from a very large book and compare the result to a much smaller book (with a size difference of a factor $n$). The figures are showing the wfd for an $n$th part (averaged over 200 sections) of the full collection and a short story by the same author. The distribution for the full collection is also included for comparison. The overlap between the short story and the section of the big book implies that the wfd for a text can be recreated by taking a section of a larger book written by the same author. It does not matter if we pull out half of a book of size $M$, or a fourth of a book of size $2M$.

These findings lead us towards \emph{the meta book concept}: The writing of a text can be described by a process where the author pulls a piece of text out of a large mother book (the meta book) and puts it down on paper. This meta book is an imaginary infinite book which gives a representation of the word frequency characteristics of everything that a certain author could ever think of writing. This has nothing to do with semantics and the actual meaning of what is written, but rather to the extent of the vocabulary, the level and type of education and the personal preferences of an author. The fact that people have such different backgrounds, together with the seemingly different behavior of the function $N(M)$ for the different authors, opens up for the speculation that every person has its own and unique meta book, in which case it can be seen as a fingerprint of an author.

Yet another, more obvious, property is the frequency of the most common word, $k_{max}$. When dividing a book in half, $k_{max}$ should also be cut in half. This linear relation between $k_{max}$ and $M$ is shown in Fig.\ 3 to be in agreement with the real data, which is consistent with the meta book concept. This follows because the most common word is most likely a ``filling word'' (e.g.\ ``the'') which would be evenly distributed throughout the text (e.g.\ every twentieth word or so).

\begin{figure}[!t]
\begin{center}
 \includegraphics[width=\columnwidth]{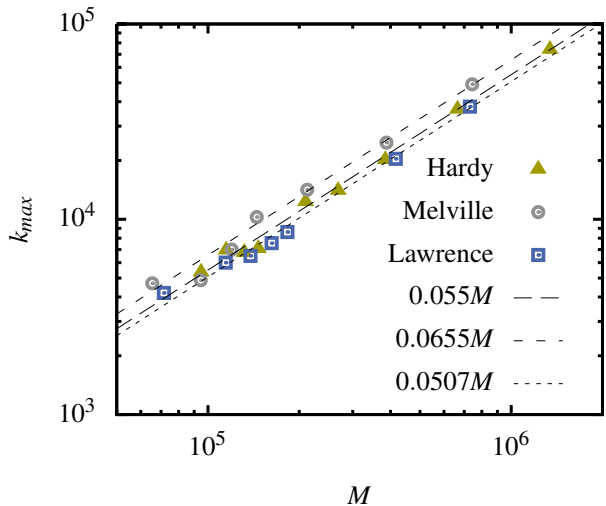}
\end{center}
\caption{The frequency of the most common word, $k_{max}$, as a function of the size of the book, $M$, for the three authors in a log-log scale.}
\end{figure}

So far we have been sectioning down books into smaller sizes, according to the meta book concept. But what happens if we go in the other direction and extrapolate to larger sizes? What could the meta book look like?
In the next section we obtain the size dependences for the parameter values of the wfd in terms of $\alpha$ and present the asymptotic limit of $\alpha = 0$.

\section{Size dependence of the wfd}
To find the size dependence of the wfd we notice that there is a simple relation between the wfd and the $\langle k \rangle$. If the $\langle k \rangle$ (which is directly related to the $N(M)$-curve, and thus to $\alpha$) is changing with the size, the wfd also has to change in some way (e.g.\ smaller cut off or changed slope). But we also know that the tail of the distribution must be regulated in such a way that the maximum frequency does not go crazy (e.g.\ 90\% of all the words are the same), but is consistent with Fig.\ 3. Given a functional form, what kind of relation between the functional parameters is needed to balance these requirements? The requirements mentioned can be summarized in three basic assumptions supported by our previous analyses: 

\begin{itemize}
\item[1] The number of different words, $N$, scales as the total number of words, $M$, to some power that can change with M, $N \propto M^{\alpha}$, where $\alpha = \alpha(M)$ can range between 1 and 0. This means that the average frequency scales like $\langle k \rangle \propto M^{1-\alpha}$.
\item[2] The value, $\tilde{k}_{max}$, defined through the cumulative word-frequency distribution as $F(\tilde{k}_{max}) = 1/N$, should increase linearly with the size of the book. That is, $\tilde{k}_{max} = \epsilon M$, where $\epsilon$ is a constant larger than zero.
\item[3] The word-frequency distribution of a book is to a good approximation of the form
\begin{equation}
P(k)=A\frac{e^{-bk}}{k^{\gamma}},
\label{func}
\end{equation}
where $A$, $b$ and $\gamma$ may depend on $M$, so that $A = A(M)$, $b = b(M) = b_0M^{-\beta}$ ($\beta \ge 0$) and $\gamma = \gamma(M)$.
\end{itemize}

The fact that $N$ can be expressed as $N \propto M^{\alpha(M)}$ is always true. The implicit assumptions made is that $\alpha(M)$ is
a slowly and monotonically decreasing function from $\alpha(1)=1$ to $\lim_{M \rightarrow \infty}\alpha(M)=0$. That a slowly varying $\alpha$ can describe $N$ is plausible since a fair approximation is usually obtained by just a constant $\alpha$ in the range $0 < \alpha < 1$ (Heap's law). The limit $\alpha(1)=1$ is just the observation that the first couple of words one writes in a book are usually different, and the limit $\alpha(M \rightarrow \infty)=0$ is the extreme limit where the author's vocabulary has been used so that no new words are added and the increase of $N$ approaches zero \cite{baayen01}.

The second assumption reflects the statement that if the most common word used by an author is ``the'' and one compares two text-lengths by the same author, where one is twice as long as the other, then the longer text contains on the average twice as many ``the''s as the shorter one. This statement can be expressed in terms of the cumulative normalized wfd, which is defined as $F(k')=\sum_{k=k'}^{\infty}P(k)$.
Thus $F(\tilde{k}_{max})=1/N$ means that if a data set is created by drawing $N$ random numbers from a theoretical and continuous function, $P(k)$, then one would get, on the average, one word appearing with a frequency larger than $\tilde{k}_{max}$. This word, with frequency $k_{max}$,  would then become the most common word in the text. So, $\tilde{k}_{max}$ is a theoretical limit, while $k_{max}$ is the actual frequency of the most common word.
Since the distribution $P(k)$ is a rapidly decreasing function for large $k$, the most common word always appear with a frequency very close to $\tilde{k}_{max}$ ($k_{max} \approx \tilde{k}_{max}$). It follows that $k_{max} \propto M$, which means that $\tilde{k}_{max} = \epsilon M$ is a vaild assumption to a good approximation.

The first two assumptions can be expressed in the continuum approximation as two integral equations:

\begin{eqnarray}
\langle k \rangle_M &=& \int_1^{\infty} kP_M(k)dk \propto M^{1-\alpha(M)} \label{int1}\\
\frac{1}{N_M} &=& \int_{\epsilon M}^{\infty} P_M(k)dk \propto M^{-\alpha(M)} \label{int2}
\end{eqnarray}

The third assumption is based on the notion that this functional form fits well to empirical data \cite{Bernhardsson09}\cite{clauset07}. The basic assumption made in the present context is that the power law with an exponential gives the correct large $k$ behavior and that $A$ and $\gamma$ vary slowly with $M$.

Next, we explore the consequences of the basic three assumptions but first the normalization condition is investigated. From Eq.\ \ref{func} we get 

\begin{eqnarray}
1 &=& \int_1^{\infty} A\frac{e^{-bk}}{k^{\gamma}}dk \approx A\int_1^{1/b} k^{-\gamma}dk\nonumber\\
&=& A\left[\frac{k^{1-\gamma}}{1-\gamma} \right]_1^{M^{\beta}/b_0} = \frac{A}{1-\gamma}\left(\frac{M^{\beta(1-\gamma)}}{b_0^{1-\gamma}}-1\right)\nonumber\\
&\Rightarrow& \textrm{for}\ M^{\beta} \gg b_0 \Rightarrow \nonumber\\
&&\left\{ \begin{array}{ll}
A \approx \gamma-1 & \textrm{if } \gamma > 1\\
A \propto M^{\beta(1-\gamma)} & \textrm{if } \gamma < 1\\
%A = \gamma-1 = const. & if\ \gamma > 1\\
%A = (1-\gamma)b_0^{1-\gamma}M^{\beta(1-\gamma)} \propto M^{1-\gamma} & if\ \gamma < 1\\
\end{array} \right.
\label{norm}
\end{eqnarray}
So, as long as $\gamma > 1$ and $M$ is sufficiently large we have no explicit $M$ dependence for $A$. That is, if $\gamma$ is constant or varies slowly enough, we can treat $A$ as constant.

The next step is to evaluate Eq.\ \ref{int1} by inserting Eq.\ \ref{func}:
\begin{eqnarray}
\langle k \rangle_M &=& \int_1^{\infty} A\frac{e^{-bk}}{k^{\gamma-1}}dk \approx A\int_1^{1/b} k^{1-\gamma}dk\nonumber\\
&=& A\left[\frac{k^{2-\gamma}}{2-\gamma} \right]_1^{M^{\beta}/b_0} = \frac{A}{2-\gamma}\left(\frac{M^{\beta(2-\gamma)}}{b_0^{2-\gamma}}-1\right)\nonumber\\
&\Rightarrow& \textrm{for}\ M^{\beta} \gg b_0 \Rightarrow \nonumber\\
&&\left\{ \begin{array}{ll}
\langle k \rangle_M = \frac{\gamma-1}{\gamma-2} & \textrm{if } \gamma > 2\\
\langle k \rangle_M \propto M^{\beta(2-\gamma)} & \textrm{if } \gamma < 2\\
%\langle k \rangle_M = \frac{\gamma-1}{(2-\gamma)b_0^{2-\gamma}}M^{\beta(2-\gamma)} \propto M^{\beta(2-\gamma)} & if\ 1 < \gamma < 2\\
%\langle k \rangle_M = \frac{\gamma-1}{\gamma-2} = const. & if\ \gamma > 2\\
\end{array} \right.
\label{kav}
\end{eqnarray}

According to Eq.\ \ref{kav} and \ref{int1} $\gamma > 2$ means that the average usage of a word is independent of the size of the book, so that $M/N = const$ and consequently $N \propto M$ ($\alpha = 1$). That is, the number of different words grows linearly with the size of the book. Solving for $\gamma$ in this case gives $\gamma=1+\frac{1}{1-1/\langle k \rangle}$. This is also the analytic solution for the Simon model \cite{simon55}, where a text grows linearly as $N = \frac{M}{\langle k \rangle}$ with preferential repetition.
Here we instead arrive at this result from the assumed functional form, without introducing any type of growth or preferential element.

However, the crucial point is that if $\gamma < 2$, then $M^{1-\alpha} \propto M^{\beta(2-\gamma)}$ and $\alpha = 1-\beta(2-\gamma)$, or

\begin{equation}
\gamma = 2-\frac{1}{\beta}(1-\alpha).
\label{gamma1}
\end{equation}

Thus, we have a relationship between $\gamma$ and $\alpha$, so the power-law exponent is determined by the rate at which new words are introduced.

The second assumption (Eq.\ \ref{int2}), with $\gamma > 1$, gives the relation

\begin{eqnarray}
 \frac{1}{N} &=& \int_{\epsilon M}^{\infty} A\frac{e^{-bk}}{k^{\gamma}}dk \approx A\int_{\epsilon M}^{1/b} k^{-\gamma}dk\nonumber\\
&=& A\left[\frac{k^{1-\gamma}}{1-\gamma} \right]_{\epsilon M}^{M^{\beta}/b_0} = \frac{A}{1-\gamma}\left(\frac{M^{\beta(1-\gamma)}}{b_0^{1-\gamma}}-(\epsilon M)^{1-\gamma}\right)\nonumber\\
&\Rightarrow& \textrm{for\ large }M \Rightarrow \left\{ \begin{array}{ll}
\frac{1}{N} \propto M^{1-\gamma} & \textrm{if } \beta \ge 1 \\
%\frac{1}{N} \propto M^{1-\gamma} & \textrm{if } \beta = 1 \\
\frac{1}{N} \propto M^{\beta(1-\gamma)} & \textrm{if } \beta < 1\\
%\frac{1}{N} = (\epsilon M)^{1-\gamma} \propto M^{1-\gamma} & if\ \beta > 1 \\
%\frac{1}{N} = M^{1-\gamma}(\epsilon^{1-\gamma}-b_0^{\gamma-1}) \propto M^{1-\gamma} & if\ \beta = 1 \\
%\frac{1}{N} = -M^{\beta(1-\gamma)}b_0^{\gamma-1} \propto M^{\beta(1-\gamma)} & if\ \beta < 1\\
\end{array} \right.
\label{1/N}
\end{eqnarray}

The last case in Eq.\ \ref{1/N} ($\beta < 1$) can be disregarded as impossible since $\gamma$ needs to be smaller than one for the integral to be positive, which means that $\alpha$ is also negative. This would give a book where the number of different words decreases as a function of the total number of words.
However, the case of $\beta \ge 1$ together with Eq.\ \ref{int2} gives the relation $1/N \propto M^{-\alpha} \propto M^{1-\gamma}$ and consequently $\alpha = \gamma-1$, or

\begin{equation}
\gamma = 1+\alpha.
\label{gamma2}
\end{equation}

Finally, substituting Eq.\ \ref{gamma2} into Eq.\ \ref{gamma1} locks down the value of $\beta$ to be one, and the wfd (given the previously assumed form) becomes:

\begin{equation}
P_M(k) = A\frac{e^{-b_0k/M}}{k^{1+\alpha(M)}},
\label{pk_final}
\end{equation}
for large $M$.

Note that if $\alpha$ goes to zero as $M$ goes to infinity, then $\gamma$ will move infinitely close to one, and this should be true \emph{for all authors}. Nevertheless, different authors might reach this point in different ways.
Taking the limit $M$ going to infinity for Eq.\ \ref{pk_final} ($b_0/M,\ \alpha(M) \rightarrow 0$) then gives us the functional form of the wfd for an infinite book:

\begin{equation}
P_{\infty}(k) = \frac{A}{k}.
\label{pk_infty}
\end{equation}
In practice though, $b_0/M$ and $\alpha(M)$ will never be exactly zero.

So far, we have shown that the meta book concept is supported by empirical data. We have also derived an expression for the size dependence of the parameters of the wfd, given a functional form. These are in some sense two independent findings which are connected through the exponent $\alpha$. 
Next we show that the derived expression for the wfd (Eq.\ \ref{pk_final}) is consistent with the real data and that the process of pulling sections out of a large book recreates the observed size dependence in $\alpha$.

\section{Size dependence in real books}
To validate the assumption that $\alpha$ approaches zero as $M$ increases, we need to fit the real data to an appropriate functional form.
This functional form needs to satisfy two constraints: $(i)$ $\alpha(M)$ should be a monotonically decreasing function with the asymptotic limit for large $M$ equal to zero; $(ii)$ $N=M^{\alpha(M)}$ should be a monotonically increasing function (by definition the number of unique words never decreases).
These constraints result in the condition
\begin{equation}
\alpha(M) \geq -M\ln M\frac{d}{dM}\alpha(M),
\label{alpha_cond}
\end{equation}
where the equality gives the solution $1/\alpha(M)=u\ln M$, where $u$ is an arbitrary constant. 
In order to parametrize $\alpha(M)$ we introduce an additional paramater, $v$, giving the expression $1/\alpha(M)=u\ln M + v$ which obeys the inequality in Eq.\ \ref{alpha_cond} if $v>0$. The final parametrization to desrcibe $\alpha$ is then

\begin{equation}
\alpha(M)=\frac{1}{u\ln M+v}.
\label{alpha_para}
\end{equation}

\begin{figure}[!t]
\begin{center}
 \includegraphics[width=\columnwidth]{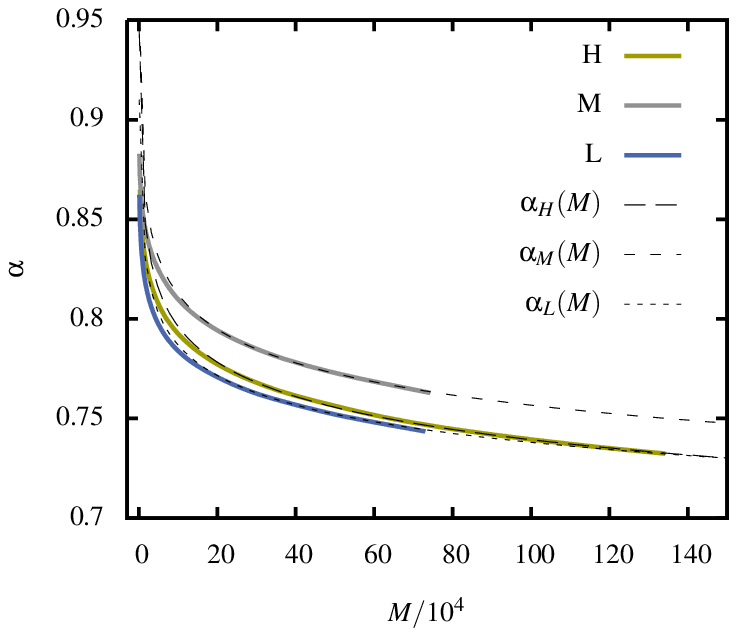}
\end{center}
\caption{The exponent $\alpha=\ln N/\ln M$ as a function of $M$ for each author together with the corresponding fits to Eq.\ \ref{alpha_para}. The fitting-parameter values ($u,v$) are for \textbf{H}ardy ($0.0420$, $0.772$), for \textbf{M}elville ($0.0394$, $0.777$) and for \textbf{L}awrence ($0.0366$, $0.849$).}
\end{figure}

The limiting value for $N$, given Eq.\ \ref{alpha_para}, is $\lim_{M \rightarrow \infty} N = \lim_{M \rightarrow \infty} M^{\alpha(M)} = e^{1/u}$. Note that this parametrization is a generalization of Heap's law ($\alpha = const$ if $u=0$).
We obtain a good fit for this parametrization for all three authors, as shown in Fig.\ 4 where we are ignoring the first $2\cdot 10^5$ words since we are interested in the large M behavior. However, the resulting fit for $N(M)=M^{\alpha(M)}$ is very resonable also for small $M$.

The main point is not to get the exact extrapolation behavior for each author but to show that they are all in accordance with the suggested functional form of $\alpha(M)$, telling us that the empirical data is consistent with $\alpha$ going to zero.

The three assumptions in the previous section lead to the specific form of the wfd in terms of $\alpha(M)$ (Eq.\ \ref{pk_final}). In Fig.\ 5 this result is compared to the real data for two authors (columns) and for each author, three different book sizes (rows). since $A$ is a  normalization constant and $\alpha(M) = \ln N/\ln M$ there is essentially only one free parameter, $b_0$.
This parameter is a characteristics of the author and according to the above analysis is independent of the length of the text. 
In other words, once the authors characteristic $b_0$ is determined then the parameter $b$ for a text of length M by the same author is given by $b=b_0/M$. The agreement suggests that the analysis leading to Eq.\ \ref{pk_final} is indeed valid.

\begin{figure}[!t]
\begin{center}
\includegraphics[width=\columnwidth]{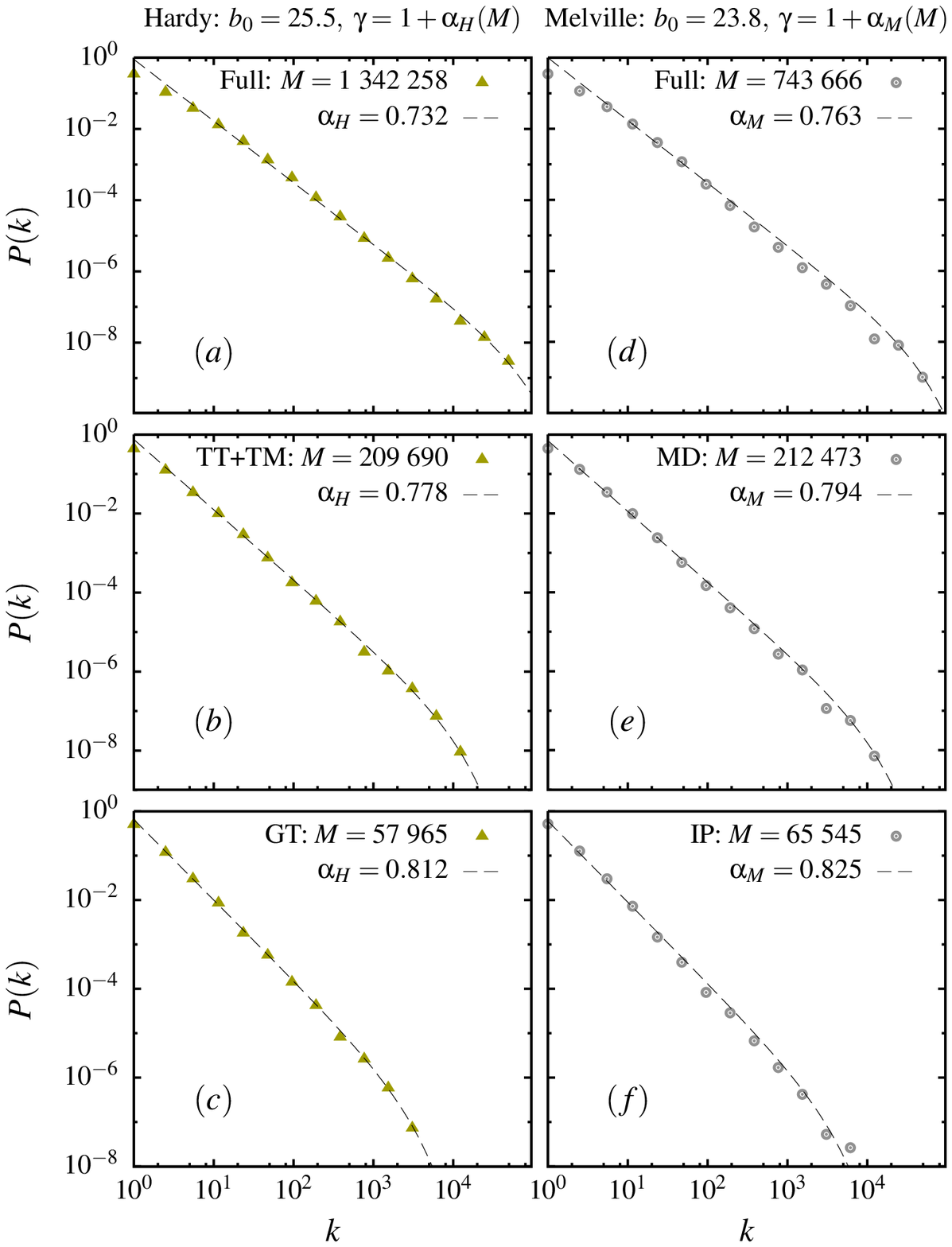}
\end{center}
\caption{The wfd for three different books (rows) of different sizes, written by Hardy (a-c) and Melville (d-f) together with the function given by Eq.\ \ref{func}. The parameters are given by $b=b_0/M$ and $\gamma=1+\alpha(M)$ (according to Eq.\ \ref{pk_final}), where $b_0=25.5$ for Hardy and $23.8$ for Melville.}
\end{figure}

The empirical data seem consistent with the size dependence derived for the wfd with $b=b_0/M$ and $\gamma = 1+\alpha(M)$. But what is causing the peculiar form of $\alpha(M)$? Our suggestion is that the actual sectioning of a book is responsible for creating such a structure. This can be tested by applying the meta book concept on a large hypothetical book.

The actual process of pulling a section out of a book can be described analytically by a combinatorial transformation, provided one assumes that the words in a book are uniformly distributed \cite{Bernhardsson09}. For instance, if the word ``the'' exists $k'$ times in a book, then the probability to get $k$ ``the'', when taking half ($n=1/2$) of that book, is given by the binomial distribution. This can be generalized for any $n$ (Eq.\ \ref{eq_rbt}) and is called the Random Book Transformation (RBT) \cite{baayen01}\cite{Bernhardsson09}.
This transformation describes how the wfd changes when a section of size $M$ is pulled out from a bigger book of size $M'$

\begin{equation}
P_M(k)=C\sum_{k^{\prime}=k}^{\infty}\boldsymbol{A}
_{kk^{\prime}}P_{M'}(k^{\prime})
\label{eq_rbt}
\end{equation}

where $n = M'/M$, $C$ is the normalization constant and $\boldsymbol{A}_{kk'}$ is the triangular matrix with the elements

\begin{equation}
%\boldsymbol{A}_{kk^{\prime}}=(n-1)^{k'-k}\frac{1}{n^{k'}}\binom{k'}{k}.\nonumber
\end{equation}

\begin{figure}[!t]
\begin{center}
 \includegraphics[width=\columnwidth]{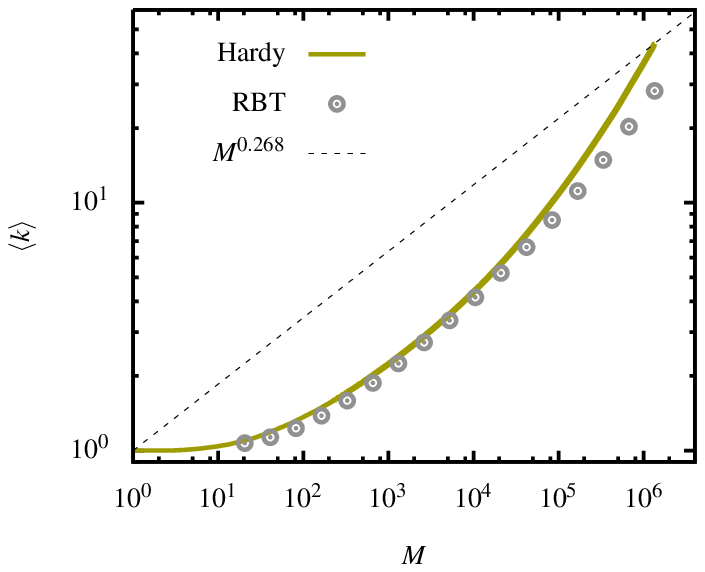}
\end{center}
\caption{The average frequency of a word, $\langle k \rangle$, as a function of the total number of words, $M$. The line shows the real data of the full collection by Hardy and the circles shows the result obtained from the RBT starting at the wfd for the full Hardy in Fig.\ 5a, i.e.\ $P_{M'}(k)=A\exp(-0.000019k)/k^{1.732}$. The dotted line correspond to the analytic solution $\langle k \rangle = M^{2-\gamma}=M^{1-\alpha}=M^{0.268}$, for a constant $\gamma=1.732$.}
\end{figure}

To analyze the behavior of the RBT we start with the theoretical wfd for the full Hardy from Fig.\ 5a ($P_{M'}(k)=A\exp(-0.000019k)/k^{1.732}$) and transform it down to smaller sizes, calculating the average frequency for each size, $M$, according to the formula

\begin{equation}
\langle k \rangle_M = \sum_{k=1}^{\infty} kP_M(k),
\label{avk_sum}
\end{equation}
where $P_M(k)$ is given by Eq.\ \ref{eq_rbt}.

In Fig.\ 6, the $\langle k \rangle_M$ is plotted in a log-log scale for the data created by the RBT, as circles, and the full line represent the real data for the full Hardy (same data as the line in Fig.\ 2a). The dotted line show the corresponding analytic result $\langle k \rangle = M^{2-\gamma}=M^{1-\alpha}=M^{0.268}$ ($\gamma = 1.732$), for a constant $\alpha$ and $\gamma$. The figure shows the similar behavior of the RBT and the real data.

\section{Conclusions}
In the present paper we have discussed the text-length dependence of the wfd of single authors. Evidence is given for a systematic decrease in the power-law index $\gamma$ of the wfd, from $\gamma \approx 2$ for short novels to the infinite book size limit with $\gamma = 1$. This systematic change is linked to the text-length dependence of the number of unique words $N$ as a function of the total number of words $M$. 

We have shown empirically that the size dependence of the wfd (and also $N$ and $\langle k \rangle$ as a function of $M$) display a very similar behavior to sectioning down a large book. It was also demonstrated, through the use of the RBT, that the same process can reproduce the observed decrease of $\alpha$. 
This has led us to introduce the concept of a meta book, which is an imaginary book of infinite length written by an author, as a description of this behavior. Furhtermore, the meta book should have a wfd close to $P(k) = A/k$.
The meta book should contain all the statistical properties of a real text, related to the specific writing style of an author, which are then transferred to the real book when pulled out of this meta book. It is important to remember that this is an abstract description, and novels (or text sections in novels), written by a single author, of length $M$ are \emph{on the average} characterized by $P_{M}(k)$. One may also note that the meta book is a holistic concept, which implies that any text length written by the author carries information about the total extent of the author's vocabulary; The $P_{M}(k)$-average for a text-section of size $M$ is independent of the total size $M'$ of the book. 

It is interesting to compare with the related phenomena of
family name distributions where the $\gamma =1$ limit is
realizable \cite{Kim05}\cite{Baek07}. In this case, $M$ corresponds to the number of
inhabitants of a country or town, $N$ to the number of different family
names, and $P(k)$ to the corresponding frequency distribution of family
names. For a country like USA or a town like Berlin $P(k)\propto k^{-\gamma}$
with $\gamma \approx 2$ \cite{newman05}\cite{Zanette01}. However, for Vietnam $\gamma
\simeq 1.4$ \cite{Baek07} and for Korea  $\gamma \simeq 1$ \cite{Kim05}. This
decrease of $\gamma$ is correlated with a corresponding decrease of $\alpha$ 
in $N \propto M^{\alpha}$. Thus the less the number of family names increases
with the size of the population, the less becomes $\gamma$, until the
limiting case $\gamma = 1$ and $\alpha = 0$ is reached. For Korea the
empirical finding is $N \propto \ln M$ , which indeed corresponds to $\alpha =0$.
In fact, the relation between the exponents $\gamma=1+\alpha$ was also achieved in Ref.\ \cite{Kim05}
for the case of family names, suggesting that the relation between $P_{M}(k)$ and $N(M)$
is more general than suggested here, and could hold for different kinds of systems.

\section{Acknowledgment}
This work was supported by the Swedish research Council through contract 50412501.

\section*{Appendix A}
\appendix

\begin{table}[]
\caption{Collection of books used as data. The authors are Thomas Hardy (TH), Herman Melville (HM) and David Herbert Lawrence (DHL).}
\begin{tabular}{l l r r}
\hline\hline
Author & Book title (abbr) & $M$ & $N$\\
\hline
TH & Greenwood Tree (GT) & 57,965 & 6,645\\[0.5ex]
    & The Well-Beloved (WB) & 63,288 & 6,985\\[0.5ex]
    & Two on a Tower (TT) & 94,849 & 8,875\\[0.5ex]
    & The Trumpet-Major (TM) & 114,841 & 9,328\\[0.5ex]
    & A Pair of Blue Eyes (BE) & 131,598 & 10,533\\[0.5ex]
    & The Woodlanders (W) & 137,184 & 10,566\\[0.5ex]
    & From the Madding Crowd (MC)& 138,004 & 11,797\\[0.5ex]
    & Desperate Remedies (DR) & 142,346 & 10,333\\[0.5ex]
    & The Hand of Ethelberta (HE) & 142,894 & 10,694\\[0.5ex]
    & Return of the Native (RN) & 142,931 & 10,437\\[0.5ex]
    & Jude the Obscure (JO) & 146,557 & 10,896\\[0.5ex]
    & Tess of the d'Urbervilles (TU) & 151,097 & 12,159\\[0.5ex]
\hline
HM & I and My Chimney (IM) & 11,525 & 2,713\\[0.5ex]
    & Israel Potter (IP) & 65,545 & 9,234\\[0.5ex]
    & The Confidence-Man (CM) & 94,644 & 10,595\\[0.5ex]
    & Typee (T) & 108,080 & 10,231\\[0.5ex]
    & Redburn. His First Voyage (RV) & 119,696 & 11,535\\[0.5ex]
    & White Jacket (WJ) & 144,892 & 13,710\\[0.5ex]
    & Moby Dick (MD) & 212,473 & 17,226\\[0.5ex]
\hline
DHL & The Prussian Officer (PO) & 9,115 & 1,823 \\[0.5ex]
      & Fantasia of the Unconscious (FU) & 61,972 & 6,192\\[0.5ex]
      & Trespasser (T) & 71,506 & 6,986\\[0.5ex]
      & Aaron's Rod (AR) & 114,384 & 8,907\\[0.5ex]
      & The Lost Girl (LG) & 137,955 & 10,427\\[0.5ex]
      & Sons and Lovers (SL) & 162,101 & 9,606 \\[0.5ex]
      & Woman in Love (WL) & 182,722 & 11,301 \\[0.5ex]
\hline
\end{tabular}
\label{book_list}
\end{table}

The empirical data used in this article are books written by three authors: Thomas Hardy, Herman Melville and David Herbert Lawrence (see table 1 for a complete list of books).
All books are taken from the online book catalog ``Project Gutenberg'' (http://www.gutenberg.org/catalog/). In order to estimate the behavior of very large books we attach together a collection of books for each author, by simply adding the books one after the other. 
Averages have been obtained by employing periodic boundary conditions and using different starting points in the book.
This is a valid procedure since the words, to large extent, are uniformly distributed throughout the book, and statistically speaking, there is no such thing as a beginning or an end \cite{Bernhardsson09}. This method gives a considerable reduction of statistical fluctuations.

When presenting the word-frequency distribution we use a $\log_2$ binning where the size of the bins follows the formula $S_i = 2^{i-1}$.

\end{document}